\documentclass[]{aastex631}

\usepackage{scrextend}

\begin{document}

\title{BSN: Photometric Light Curve Analysis of Two Contact Binary Systems LS Del and V997 Cyg}

\author[0000-0002-0196-9732]{Atila Poro}
\altaffiliation{poroatila@gmail.com}
\affiliation{Astronomy Department of the Raderon AI Lab., BC., Burnaby, Canada}

\author{Mehmet Tanriver}
\affiliation{Department of Astronomy and Space Science, Faculty of Science, Erciyes University, Kayseri TR-38039, Türkiye}
\affiliation{Erciyes University, Astronomy and Space Science Observatory Application and Research Center, Kayseri TR-38039, Türkiye}

\author{Elham Sarvari}
\affiliation{Independent Astrophysics Researcher, Tehran, Iran}

\author{Shayan Zavvarei}
\affiliation{Department of Physics, Shahid Beheshti University, Tehran, Iran}

\author{Hossein Azarara}
\affiliation{Faculty of Physics, Shahid Bahonar University of Kerman, P.O.Box 76175, Kerman, Iran}

\author{Laurent Corp}
\affiliation{Double Stars Committee, Société Astronomique de France, Paris, France}

\author{Sabrina Baudart}
\affiliation{Double Stars Committee, Société Astronomique de France, Paris, France}

\author{Asma Ababafi}
\affiliation{Dez Astronomical Association, Khouzestan Province, Dezful, Iran}

\author{Nazanin Kahali Poor}
\affiliation{Independent Astrophysics Researcher, Tehran, Iran}

\author{Fariba Zare}
\affiliation{Department of Physics, Amirkabir University of Technology, Tehran, Iran}

\author{Ahmet Bulut}
\affiliation{Department of Physics, Faculty of Sciences, Çanakkale Onsekiz Mart University, Terzioğlu Kampüsü, TR-17020, Çanakkale, Türkiye}
\affiliation{Astrophysics Research Center and Observatory, Çanakkale Onsekiz Mart University, Terzioğlu Kampüsü, TR-17020, Çanakkale, Türkiye}

\author{Ahmet Keskin}
\affiliation{Department of Astronomy and Space Science, Faculty of Science, Erciyes University, Kayseri TR-38039, Türkiye}

\begin{abstract}
The light curve analyses and orbital period variations for two contact binary stars, LS Del and V997 Cyg, were presented in this work which was conducted in the frame of the Binary Systems of South and North (BSN) Project. Ground-based photometric observations were performed at two observatories in France. We used the TESS (Transiting Exoplanet Survey Satellite) data for extracting times of minima and light curve analysis of the target systems. The O-C diagram for both systems displays a parabolic trend. LS Del and V997 Cyg's orbital periods are increasing at a rate of $dP/dt= 7.20\times10^{-08}$ $d\ yr^{-1}$ and $dP/dt= 2.54\times10^{-08}$ $d\ yr^{-1}$, respectively. Therefore, it can be concluded that the mass is being transferred from the less massive star to the more massive component with a rate of $dM/dt=-1.96\times10^{-7}$ $M_\odot$ $yr^{-1}$ for the LS Del system, and $dM/dt=-3.83\times10^{-7}$ $M_\odot$ $yr^{-1}$ for V997 Cyg. The parameters of the third possible object in the system were also considered. The PHysics Of Eclipsing BinariEs (PHOEBE) Python code was used to analyze the light curves. The light curve solutions needed a cold starspot due to the asymmetry in the LS Del system's light curve maxima. The mass ratio, fill-out factor, and star temperature all indicate that both systems are contact binary types in this investigation. Two methods were used to estimate the absolute parameters of the systems: one method used the parallax of Gaia DR3, and the other used a $P-M$ relationship. The positions of the systems were also depicted on the $M-L$, $M-R$, $q-L_{ratio}$, and $logM_{tot}-logJ_0$ diagrams. We recommended that further observations and investigations be done on the existence of the fourth body in this system.
\end{abstract}


\section{Introduction}
W Ursae Majoris (W UMa) type systems are categorized as eclipsing variables, with both components fulfilling their Roche lobes, as described by \cite{1971ARA&A...9..183P}. A common convective envelope is shared between the two components (\citealt{1968ApJ...151.1123L}), in which mass and energy transfer from one component to the other. Consequently, despite having different features, the two components of contact binaries maintain nearly identical temperatures.

The orbital period of contact systems is typically $P_{orb} \leq 1$ day, with most systems having a period of $\sim 0.2-0.6$ days (\citealt{2003CoSka..33...38P}, \citealt{2006Ap.....49..358D}, \citealt{2021ApJS..254...10L}). Also, various studies have shown that some of the absolute parameters of stars in contact systems can have a significant relationship with the orbital period (\citealt{2003MNRAS.342.1260Q}, \citealt{2008MNRAS.390.1577G}, \citealt{2018PASJ...70...90K}). Additionally, it is known that contact binaries show significant magnetic activity, which affects orbital period fluctuations (\citealt{1951PRCO....2...85O}, \citealt{1992ApJ...385..621A}).

Despite extensive studies conducted on W UMa-type systems, however, there are still many open questions regarding the physical and geometric features of the components and the formation and evolution of the system (\citealt{2016ApJ...817..133Z}).
\\
\\
We investigated two binary star systems. The first target in this study is LS Delphini (GSC 01656-01961). It was discovered by \cite{1976IBVS.1214....1B}, who used the $y$ filter in photoelectric observations.
Observations showed that LS Del has a magnitude ranging from 8.66 to 9.17 in the $V$ filter and an orbital period of 0.363842 days reported in the VSX database.
Studies have been conducted regarding the light curve analysis of LS Del such as \cite{1996AA...311..523M}, \cite{1999AJ....118..515L}, and \cite{2011MNRAS.412.1787D}. In all of the studies, the inclination was estimated less than 50 degrees. Also, a noteworthy point in the analysis of the photometric light curve and with spectroscopic data is that this system's mass ratio is reported to be between 0.37 and 0.57.

The second system, V997 Cyg (GSC 03935-02233), is in the Cygnus constellation that was discovered by Cuno Hoffmeister (\citealt{1963AN....287..169H}). It has a short orbital period of less than half a day (0.4582264\footnote{ASAS-SN Catalog of Variable Stars \url{http://asas-sn.osu.edu/variables}} days), and its magnitude range in the $V$ filter is between 13.16 and 13.61\footnote{The International Variable Star Index (VSX) \url{https://www.aavso.org/vsx}}.
Initially, V997 Cyg was categorized as an RR Lyr type in the Sonneberg Obs. catalog (\citealt{1966VeSon...7...61G}), and it was later recognized as an eclipsing binary by the \cite{2000AJ....119.1901A} study. There were observations related to the V997 Cyg system, such as \cite{2008AJ....135..850D}.
\\
\\
This work is a continuation of the BSN\footnote{\url{https://bsnp.info/}} project's observations and analyses of eclipsing binary systems. Therefore, we employed ground-based observations and TESS data for studying the LS Del and V997 Cyg binary systems. We presented an analysis of the orbital period variations and a new ephemeris for each target system. Light curve solutions for both target systems were provided; however, the analysis of the V997 Cyg system was done for the first time. Additionally, two methods were used to estimate the systems' absolute parameters.

\vspace{1.5cm}
\section{Observation and Data Reduction}
The LS Del and V997 Cyg binary systems were observed in two observatories in France. The $V$ standard filter has been employed for observations at both observatories.
\\
\\
LS Del was observed on five nights during the summer of 2016, 2019, 2021, and 2023 at a private observatory (long 02 29 09 E, lat 44 15 38 N, alt of 661m). We employed a 135mm Telelens and a CCD Moravian G2 for these observations. We had a total of 956 images with 60 seconds of exposure time. The average temperature of the CCD was $-25^{\circ}$C during the observations.
Basic data reduction was carried out using Maxim DL 5.24 software (\citealt{2000IAPPP..79....2G}), and the bias, dark, and flat-field images followed the standard technique. We used SAO 106677 (RA.: 314.0342, Dec.: 19.8394) and SAO 89337 (RA.: 314.6170 and Dec.: 20.3980) as comparison stars. SAO 106663 (RA.: 313.8770 and Dec.: 19.6862) was also considered to be a check star.
Also, the maximum apparent magnitude of the system's light curve was $V_{max}=8.62(6){mag}$.
\\
\\
V997 Cyg was observed in a private observatory in Toulon, France, at a longitude of $05^{\circ}$ $54'$ $35"$ E and a latitude of $43^{\circ}$ $8'$ $59"$ N, and an altitude of 68 meters above mean sea level. The observations were carried out on four nights from June to July 2023.
Observations were employed an Apochromatic Refractor TS Optics with 102mm aperture and a ZWO ASI 1600MM CCD. The binning of the images was $1\times1$, with a 180-second exposure time and the average temperature of the CCD was $-15^{\circ}$C.
The basic data reduction was carried out by the bias, dark, and flat fields image and with the Siril 1.2.0-rc2 program.
We used Gaia DR2 2137381511262134016 (RA.: 296.9277 and Dec.: 52.8988) and Gaia DR2 2137287331220260096 (RA.: 296.8882 and Dec.: 52.8103) as comparison and check stars, respectively.
According to our observations, the light curve's maximum apparent magnitude was obtained to be $V_{max}=13.19(9){mag}$.
\\
\\
TESS observed the LS Del (TIC 265972568) and V997 Cyg (TIC 27997238) binary systems. We used sector 55 with an exposure time of 120 seconds for the LS Del target system.
For V997 Cyg, We used sectors 14 (1800-second exposure time), 15 (1800-second exposure time), 16 (1800-second exposure time), 41 (600-second exposure time), 54 (600-second exposure time), 55 (600-second exposure time), and 56 (200-second exposure time).
The data is available at the Mikulski Space Telescope Archive (MAST)\footnote{\url{https://mast.stsci.edu/portal/Mashup/Clients/Mast/Portal.htmL}}.

\vspace{1.5cm}
\section{Orbital Period Variations}
According to the ground-based observations for this study, we derived the primary and secondary times of minima. So, one primary and three secondary minima were extracted for the LS Del system, whereas for the V997 Cyg system, two primary and one secondary minima were extracted. Then, we performed an exhaustive search to find all times of minima reported in studies, databases, and catalogs. We obtained the data from the TESS observations and the AAVSO (The American Association of Variable Star Observers), ASAS (The All Sky Automated Survey), ASAS-SN (The All Sky Automated Survey for SuperNovae), ZTF (The Zwicky Transient Facility), TrES (The Trans-Atlantic Exoplanet Survey), SWASP (The Super Wide Angle Search for Planets), and QES (The Qatar Exoplanet Survey) for both systems, and we used them to extract times of minima (Appendix Tables \ref{tabA1} to \ref{tabA7}). The timing data used in the O-C analysis of LS Del and V997 Cyg covers almost 48 and 61 years, respectively.
Then, using the reference ephemeris (Equations \ref{eq1} and \ref{eq2}), we calculated the epoch and O-C values.

\begin{equation}\label{eq1}
\left\{\begin{array}{l}
LS\ Del:\\
Min.I(BJD_{TDB})=2442687.4185+0.3638404\times E\\
Bond\ (1976)\\
\end{array}\right.
\end{equation}

\begin{equation}\label{eq2}
\left\{\begin{array}{l}
V997\ Cyg:\\
Min.I(BJD_{TDB})=2455460.5143+0.458226\times E\\
Arena\ et\ al.\ (2011)\\
\end{array}\right.
\end{equation}

\begin{figure*}
\begin{center}
\includegraphics[width=\textwidth]{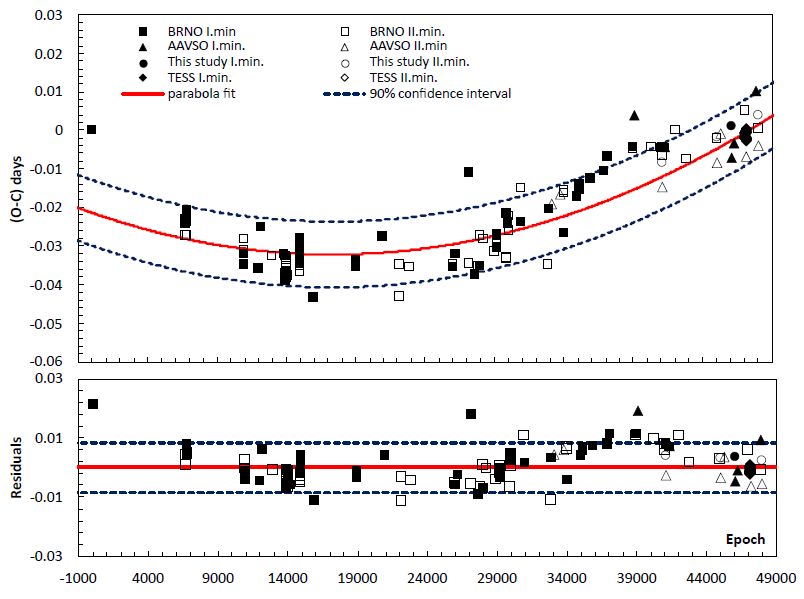}
\caption{The O-C diagram of LS Del with the parabolic fit to the O-C dataset is shown with a solid red line. Residuals from the parabolic fit are shown in the bottom part of the diagram.}
\label{Fig1}
\end{center}
\end{figure*}

\begin{figure*}
\begin{center}
\includegraphics[width=\textwidth]{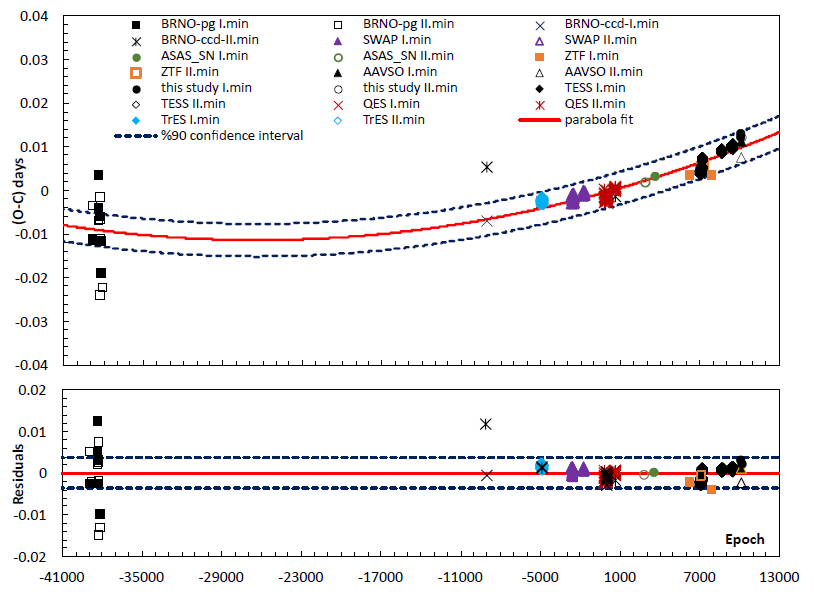}
\caption{The V997 Cyg system's O-C diagram. The solid red line showed the parabolic fit to the O-C dataset. The diagram's bottom displays residuals of the fit.}
\label{Fig2}
\end{center}
\end{figure*}

The O-C diagrams of the LS Del and V997 Cyg systems were presented in Figures \ref{Fig1} and \ref{Fig2} and the
residuals of the fitting. As shown in the O-C diagrams of the target systems, both have an upward parabolic trend. We computed a new ephemeris for each of the systems (Equations \ref{eq3} and \ref{eq4}):

\begin{equation}\label{eq3}
\left\{\begin{array}{l}
LS\ Del:\\
Min.I(BJD_{TDB})=2442687.3508(16)+0.363841821(37)\times E\\
\end{array}\right.
\end{equation}

\begin{equation}\label{eq4}
\left\{\begin{array}{l}
V997\ Cyg:\\
Min.I(BJD_{TDB})=2455460.5144(1)+0.458226914(17)\times E\\
\end{array}\right.
\end{equation}

where $E$ is the cycle number after the reference cycle. Both target systems analyzed for orbital period changes in this work have observations over a long time. The O-C parabolic variations suggest the orbital periods of LS Del and V997 Cyg systems are increasing at a rate of $dP/dt= 7.20\times10^{-08}$ $d\ yr^{-1}$ and $dP/dt= 2.54\times10^{-08}$ $d\ yr^{-1}$, respectively.

Previous studies on the LS Del system have similarly indicated a tendency of changes with parabolic fit.
The results of the \cite{1991ApSS.186...57D} study displayed an increasing parabolic fit and the O-C diagram indicates the mass transfer from a less massive star to a more massive one. The \cite{2001MNRAS.328..635Q} study also fit a parabolic trend on the O-C diagram of LS Del and presented a continuous period increase rate of $dP/dt=2.25\times10^{-07}$ $d\ yr^{-1}$.

We analyzed the V997 Cyg system for the first time and the parabolic fit on the O-C diagram can be considered based on the data. The parabolic fit for the LS Del and V997 Cyg systems could show star mass transfer in contact binary systems.

\begin{table*}
\caption{Available times of minima for the LS Del system from the literature and this study.}
\renewcommand{\arraystretch}{0.8}
\centering
\begin{center}
\footnotesize
\begin{tabular}{c c c c c c c c}
 \hline
 \hline
Min.($BJD_{TDB}$) & Epoch & O-C & Reference & Min.($BJD_{TDB}$) & Epoch & O-C & Reference\\
\hline
2442687.4185	&	0	&	0	&	\cite{1976IBVS.1214....1B}	&	2460081.5467	&	47807	&	0.0102	&	AAVSO	\\
2445136.4051	&	6731	&	-0.0232	&	\cite{1984IBVS.2553....1S}	&	2460139.5651	&	47966.5	&	-0.0040	&	AAVSO	\\
2445145.5001	&	6756	&	-0.0242	&	\cite{1984IBVS.2553....1S}	&	2447778.4186(16)	&	13992.5	&	-0.0367	&	\cite{1989IBVS.3406....1W}	\\
2445146.4064	&	6758.5	&	-0.0275	&	\cite{1984IBVS.2553....1S}	&	2449588.3433(8)	&	18967	&	-0.0356	&	\cite{1994IBVS.4126....1D}	\\
2445149.5059	&	6767	&	-0.0206	&	\cite{1984IBVS.2553....1S}	&	2449588.3458(7)	&	18967	&	-0.0336	&	\cite{1994IBVS.4126....1D}	\\
2445150.4088	&	6769.5	&	-0.0273	&	\cite{1984IBVS.2553....1S}	&	2450301.4790(15)	&	20927	&	-0.0276	&	\cite{1999IBVS.4670....1S}	\\
2445177.3363	&	6843.5	&	-0.0240	&	\cite{1984IBVS.2553....1S}	&	2450731.3408(17)	&	22108.5	&	-0.0432	&	\cite{1999IBVS.4670....1S}	\\
2445177.5190	&	6844	&	-0.0232	&	\cite{1984IBVS.2553....1S}	&	2450758.2733(9)	&	22182.5	&	-0.0349	&	\cite{1999IBVS.4670....1S}	\\
2446668.1615	&	10941	&	-0.0348	&	\cite{1987IBVS.2982....1W}	&	2452134.4990(2)	&	25965	&	-0.0355	&	\cite{2005IBVS.5623....1D}	\\
2446670.1692	&	10946.5	&	-0.0283	&	\cite{1987IBVS.2982....1W}	&	2452136.5007(2)	&	25970.5	&	-0.0349	&	\cite{2005IBVS.5623....1D}	\\
2446671.0750	&	10949	&	-0.0321	&	\cite{1987IBVS.2982....1W}	&	2452200.3577(2)	&	26146	&	-0.0320	&	\cite{2002IBVS.5313....1B}	\\
2447028.3626	&	11931	&	-0.0357	&	\cite{1989IBVS.3406....1W}	&	2452550.5516(2)	&	27108.5	&	-0.0345	&	\cite{2003IBVS.5378....1D}	\\
2447114.2394	&	12167	&	-0.0253	&	\cite{1991AAS...90..301D}	&	2452562.4001(3)	&	27141	&	-0.0108	&	\cite{2009IBVS.5887....1Y}	\\
2447386.5666	&	12915.5	&	-0.0326	&	\cite{1991AAS...90..301D}	&	2452808.5138(10)	&	27817.5	&	-0.0351	&	\cite{2004IBVS.5579....1B}	\\
2447729.4866	&	13858	&	-0.0322	&	\cite{1989IBVS.3406....1W}	&	2452854.9032(11)	&	27945	&	-0.0353	&	\cite{2004IBVS.5494....1M}	\\
2447737.4906	&	13880	&	-0.0326	&	\cite{1989IBVS.3406....1W}	&	2452855.0932(21)	&	27945.5	&	-0.0273	&	\cite{2004IBVS.5494....1M}	\\
2447741.4865	&	13891	&	-0.0390	&	\cite{1991AAS...90..301D}	&	2452952.6013(6)	&	28213.5	&	-0.0284	&	\cite{2004IBVS.5502....1D}	\\
2447745.4906	&	13902	&	-0.0371	&	\cite{1991AAS...90..301D}	&	2453229.4808(9)	&	28974.5	&	-0.0314	&	\cite{2006IBVS.5684....1B}	\\
2447772.4153	&	13976	&	-0.0366	&	\cite{1991AAS...90..301D}	&	2453293.3390(3)	&	29150	&	-0.0272	&	\cite{2009IBVS.5887....1Y}	\\
2447778.4216	&	13992.5	&	-0.0337	&	\cite{1989IBVS.3406....1W}	&	2453302.2534(5)	&	29174.5	&	-0.0269	&	\cite{2005IBVS.5649....1A}	\\
2447790.4236	&	14025.5	&	-0.0384	&	\cite{1989IBVS.3406....1W}	&	2453303.3426(7)	&	29177.5	&	-0.0292	&	\cite{2005IBVS.5649....1A}	\\
2447790.4246	&	14025.5	&	-0.0374	&	\cite{2001MNRAS.328..635Q}	&	2453304.2509(6)	&	29180	&	-0.0305	&	\cite{2005IBVS.5649....1A}	\\
2447790.4247	&	14025.5	&	-0.0373	&	\cite{1991AAS...90..301D}	&	2453558.3909(5)	&	29878.5	&	-0.0331	&	\cite{2009IBVS.5887....1Y}	\\
2447790.4253	&	14025.5	&	-0.0367	&	\cite{1991AAS...90..301D}	&	2453559.4822(4)	&	29881.5	&	-0.0333	&	\cite{2006IBVS.5684....1B}	\\
2447790.4276	&	14025.5	&	-0.0345	&	\cite{1991ApSS.186...57D}	&	2453560.4034(4)	&	29884	&	-0.0217	&	\cite{2009IBVS.5887....1Y}	\\
2447790.4290	&	14025.5	&	-0.0330	&	\cite{kreiner2001atlas}\footnote{\url{http://www.as.wsp.krakow.pl/o-c}}	&	2453589.3266(6)	&	29963.5	&	-0.0238	&	\cite{2009IBVS.5887....1Y}	\\
2447822.2602	&	14113	&	-0.0379	&	\cite{1991AAS...90..301D}	&	2453589.5082(9)	&	29964	&	-0.0241	&	\cite{2009IBVS.5887....1Y}	\\
2448119.5214	&	14930	&	-0.0343	&	\cite{1991ApSS.186...57D}	&	2453606.4247(8)	&	30010.5	&	-0.0262	&	\cite{2009IBVS.5887....1Y}	\\
2448119.5277	&	14930	&	-0.0280	&	\cite{1991ApSS.186...57D}	&	2453613.3417(4)	&	30029.5	&	-0.0222	&	\cite{2005IBVS.5649....1A}	\\
2448122.4337	&	14938	&	-0.0328	&	\cite{1991ApSS.186...57D}	&	2453937.5308(3)	&	30920.5	&	-0.0149	&	\cite{2007IBVS.5753....1B}	\\
2448122.4358	&	14938	&	-0.0307	&	\cite{1991ApSS.186...57D}	&	2453938.4313(3)	&	30923	&	-0.0240	&	\cite{2007IBVS.5753....1B}	\\
2448123.3392	&	14940.5	&	-0.0369	&	\cite{1991ApSS.186...57D}	&	2454628.8075(10)	&	32820.5	&	-0.0348	&	\cite{2011MNRAS.412.1787D}	\\
2448123.3409	&	14940.5	&	-0.0352	&	\cite{1991ApSS.186...57D}	&	2454650.4706(5)	&	32880	&	-0.0202	&	\cite{2009IBVS.5898....1P}	\\
2448129.3476	&	14957	&	-0.0318	&	\cite{1991ApSS.186...57D}	&	2455050.5171(7)	&	33979.5	&	-0.0163	&	\cite{2010IBVS.5941....1H}	\\
2448129.3492	&	14957	&	-0.0302	&	\cite{1991ApSS.186...57D}	&	2455059.4209(6)	&	34004	&	-0.0265	&	\cite{2010IBVS.5941....1H}	\\
2448131.3438	&	14962.5	&	-0.0368	&	\cite{1991ApSS.186...57D}	&	2455059.6138(7)	&	34004.5	&	-0.0156	&	\cite{2010IBVS.5941....1H}	\\
2448131.3464	&	14962.5	&	-0.0342	&	\cite{1991ApSS.186...57D}	&	2455401.4402(8)	&	34944	&	-0.0172	&	\cite{2011IBVS.5980....1P}	\\
2448472.4375	&	15900	&	-0.0434	&	\cite{1994IBVS.4027....1M}	&	2455463.2950(8)	&	35114	&	-0.0153	&	\cite{2011IBVS.5980....1P}	\\
2449588.3438	&	18967	&	-0.0356	&	\cite{1994IBVS.4127....1H}	&	2455476.3944(6)	&	35150	&	-0.0141	&	\cite{2011IBVS.5980....1P}	\\
2449588.3458	&	18967	&	-0.0336	&	\cite{1994IBVS.4127....1H}	&	2455735.4505(3)	&	35862	&	-0.0125	&	\cite{2013IBVS.6044....1P}	\\
2451000.2264	&	22847.5	&	-0.0357	&	\cite{1999IBVS.4681....1K}	&	2456105.4780(12)	&	36879	&	-0.0107	&	\cite{2013IBVS.6044....1P}	\\
2452734.8338	&	27615	&	-0.0374	&	ASAS	&	2456105.4781(6)	&	36879	&	-0.0106	&	\cite{2013IBVS.6044....1P}	\\
2454722.6942	&	33078.5	&	-0.0190	&	AAVSO	&	2456180.4329(4)	&	37085	&	-0.0069	&	\cite{2013IBVS.6044....1P}	\\
2454947.9139	&	33697.5	&	-0.0165	&	AAVSO	&	2456852.8127(7)	&	38933	&	-0.0041	&	\cite{2015IBVS.6131....1N}	\\
2456898.6645	&	39059	&	0.0038	&	AAVSO	&	2456866.8198(7)	&	38971.5	&	-0.0049	&	\cite{2015IBVS.6131....1N}	\\
2457608.5087	&	41010	&	-0.0046	&	BRNO	&	2457329.6255(3)	&	40243.5	&	-0.0042	&	\cite{2016JAVSO..44...69S}	\\
2457613.4106	&	41023.5	&	-0.0146	&	AAVSO	&	2457613.4169(4)	&	41023.5	&	-0.0082	&	This study	\\
2457636.7068	&	41087.5	&	-0.0042	&	\cite{2017JAVSO..45..121S}	&	2457974.3548(4)	&	42015.5	&	0.0000	&	\cite{2019OEJV..203....1O}	\\
2457704.5625	&	41274	&	-0.0047	&	BRNO	&	2458262.5091(1)	&	42807.5	&	-0.0073	&	\cite{2019OEJV..203....1O}	\\
2457704.5628	&	41274	&	-0.0044	&	AAVSO	&	2459049.5012(9)	&	44970.5	&	-0.0021	&	\cite{2020JAVSO..48..256S}	\\
2459146.6477	&	45237.5	&	-0.0009	&	AAVSO	&	2459049.5016(5)	&	44970.5	&	-0.0017	&	This study	\\
2459502.6631	&	46216	&	-0.0033	&	AAVSO	&	2459440.4508(3)	&	46045	&	0.0011	&	This study	\\
2459784.4660	&	46990.5	&	0.0051	&	AAVSO	&	2460139.5730(4)	&	47966.5	&	0.0040	&	This study	\\
2459826.6594	&	47106.5	&	-0.0069	&	AAVSO	&		&		&		&		\\
\hline
\hline
\end{tabular}
\end{center}
\label{tab1}
\end{table*}

\begin{table*}
\caption{Available times of minima for the V997 Cyg system from the literature and this study.}
\centering
\begin{center}
\footnotesize
\begin{tabular}{c c c c c c c c}
 \hline
 \hline
Min.($BJD_{TDB}$) & Epoch & O-C & Reference & Min.($BJD_{TDB}$) & Epoch & O-C & Reference\\
\hline
2437668.2734	&	-38828.5	&	-0.0126	&	BRNO	&	2455462.3453(7)	&	4	&	-0.0019	&	\cite{2011IBVS.5997....1A}	\\
2437669.3934	&	-38826	&	-0.0382	&	BRNO	&	2455463.4925(12)	&	6.5	&	-0.0003	&	\cite{2011IBVS.5997....1A}	\\
2437696.6584	&	-38766.5	&	-0.0376	&	BRNO	&	2455469.4482(4)	&	19.5	&	-0.0015	&	\cite{2011IBVS.5997....1A}	\\
2437903.5714	&	-38315	&	-0.0137	&	BRNO	&	2455469.4489(17)	&	19.5	&	-0.0008	&	\cite{2011IBVS.5997....1A}	\\
2437904.5124	&	-38313	&	0.0109	&	BRNO	&	2455472.4251(20)	&	26	&	-0.0031	&	\cite{2011IBVS.5997....1A}	\\
2437906.5404	&	-38308.5	&	-0.0232	&	BRNO	&	2455476.3215(22)	&	34.5	&	-0.0016	&	\cite{2011IBVS.5997....1A}	\\
2437933.5614	&	-38249.5	&	-0.0375	&	BRNO	&	2455478.3844(11)	&	39	&	-0.0007	&	\cite{2011IBVS.5997....1A}	\\
2437934.5094	&	-38247.5	&	-0.0059	&	BRNO	&	2455479.3002(4)	&	41	&	-0.0014	&	\cite{2011IBVS.5997....1A}	\\
2437939.5334	&	-38236.5	&	-0.0224	&	BRNO	&	2455764.5457(5)	&	663.5	&	-0.0015	&	\cite{2012IBVS.6033....1B}	\\
2437940.4334 & -38234.5	&	-0.0389	&	BRNO	&	2456785.0182 & 2890.5	&	0.0017	&	ASAS-SN	\\
2437944.5154 & -38225.5	&	-0.0809	&	BRNO	&	2457121.1283 & 3624	&	0.0030	&	ASAS-SN	\\
2437947.5544	&	-38219	&	-0.0204	&	BRNO	&	2458336.8021	&	6277	&	0.0032	&	ZTF	\\
2437959.4494	&	-38193	&	-0.0393	&	BRNO	&	2458723.7764	&	7121.5	&	0.0057	&	ZTF	\\
2437970.4224	&	-38169	&	-0.0637	&	BRNO	&	2458763.6423	&	7208.5	&	0.0060	&	ZTF	\\
2438001.3414	&	-38101.5	&	-0.0749	&	BRNO	&	2459082.7941	&	7905	&	0.0033	&	ZTF	\\
2451291.8085(22)	&	-9097.5	&	0.0053	&	\cite{2001IBVS.5027....1D}	&	2460118.3926	&	10165	&	0.0111	&	AAVSO	\\
2451304.8555(27)	&	-9069	&	-0.0071	&	\cite{2001IBVS.5027....1D}	&	2460118.3945(7)	&	10165	&	0.0129	&	This study	\\
2453210.1638	&	-4911	&	-0.0026	&	BRNO	&	2460118.6182	&	10165.5	&	0.0075	&	AAVSO	\\
2453210.3928	&	-4910.5	&	-0.0027	&	BRNO	&	2460123.4340(20)	&	10176	&	0.0120	&	This study	\\
2455459.3675(14)	&	-2.5	&	-0.0012	&	\cite{2011IBVS.5997....1A}	&	2460130.5365(8)	&	10191.5	&	0.0120	&	This study	\\
2455460.5143(17)	&	0	&	0	&	\cite{2011IBVS.5997....1A}	&		&		&		&		\\
\hline
\hline
\end{tabular}
\end{center}
\label{tab2}
\end{table*}

\vspace{1.5cm}
\section{Light Curve Solutions}
The LS Del and V997 Cyg systems photometric light curve solutions were carried out using the PHOEBE Python code version 2.4.9 (\citealt{2016ApJS..227...29P}, \citealt{2020ApJS..250...34C}).

We assumed the bolometric albedo and gravity-darkening coefficients were $A_1=A_2=0.5$ (\citealt{1969AcA....19..245R}) and $g_1=g_2=0.32$ (\citealt{1967ZA.....65...89L}), respectively.
We modeled the stellar atmosphere using the \cite{2004AA...419..725C} method and the limb darkening coefficients were adopted as a free parameter in the PHOEBE code. The reflection effect was also taken into account in our contact binary systems (e.g. \citealt{1990ApJ...356..613W}, \citealt{2016ApJS..227...29P}).

It should be mentioned that, regarding the use of TESS data for both systems, we only used one day of observations for each system. It appears that not all TESS data for all observation days can be used to provide an acceptable light curve analysis since the light curves will be thick. Thus, for LS Del and V997 Cyg, we utilized TESS data in sectors 55 and 56, respectively.

The initial system effective temperature from Gaia DR3 was taken into consideration and we set it on the hotter component. We used the depth difference between the light curve minima to estimate the temperature ratio.
We set $q=0.375$ for the LS Del system as an entitled value for light curve analysis from the \cite{1999AJ....118..515L} and \cite{2011MNRAS.412.1787D} studies. Also, considering that photometric data was available, the mass ratio of V997 Cyg was estimated using $q$-search. Then, following obtaining the temperatures, setting the mass ratio, the estimation of the inclination, and the fillout factor, we attempted to obtain a good synthetic light curve.

Asymmetric maxima in the light curves of contact binary stars could be an indication of the well-known O'Connell effect (\citealt{1951PRCO....2...85O}) and the presence of magnetic activity. The asymmetry in the LS Del system's light curve maxima indicates that a cold spot on the secondary star was required for the light curve's solution. It should be noted that the light curve analysis of the V997 Cyg system was performed without a starspot.

\begin{figure*}
\begin{center}
\includegraphics[width=\textwidth]{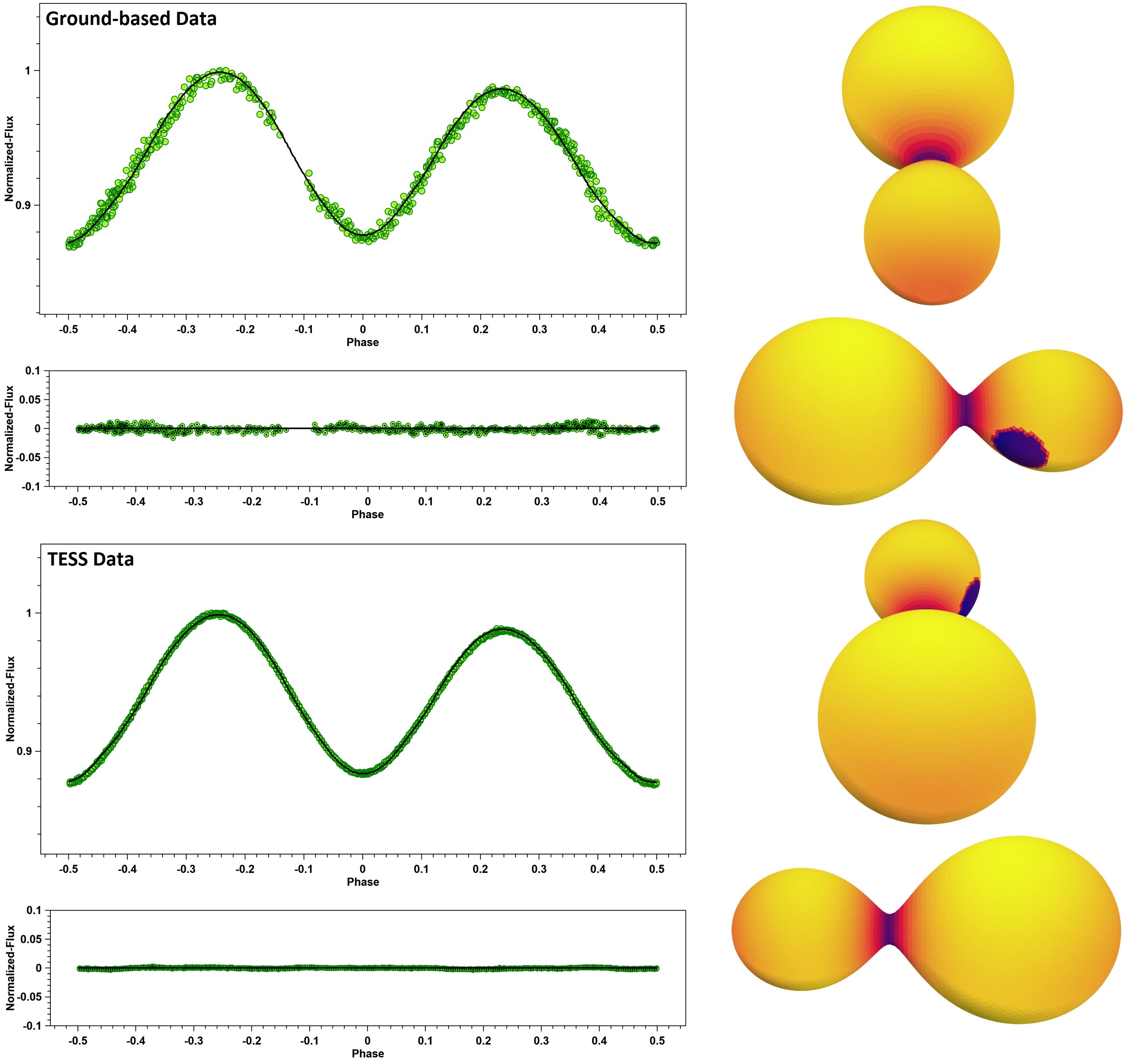}
\caption{Observational (TESS and ground-based) and synthetic light curves, along with the three-dimensional view in four phases of the LS Del system. The green points are the data and the black line is the synthetic fit.}
\label{Fig3}
\end{center}
\end{figure*}

\begin{figure*}
\begin{center}
\includegraphics[width=\textwidth]{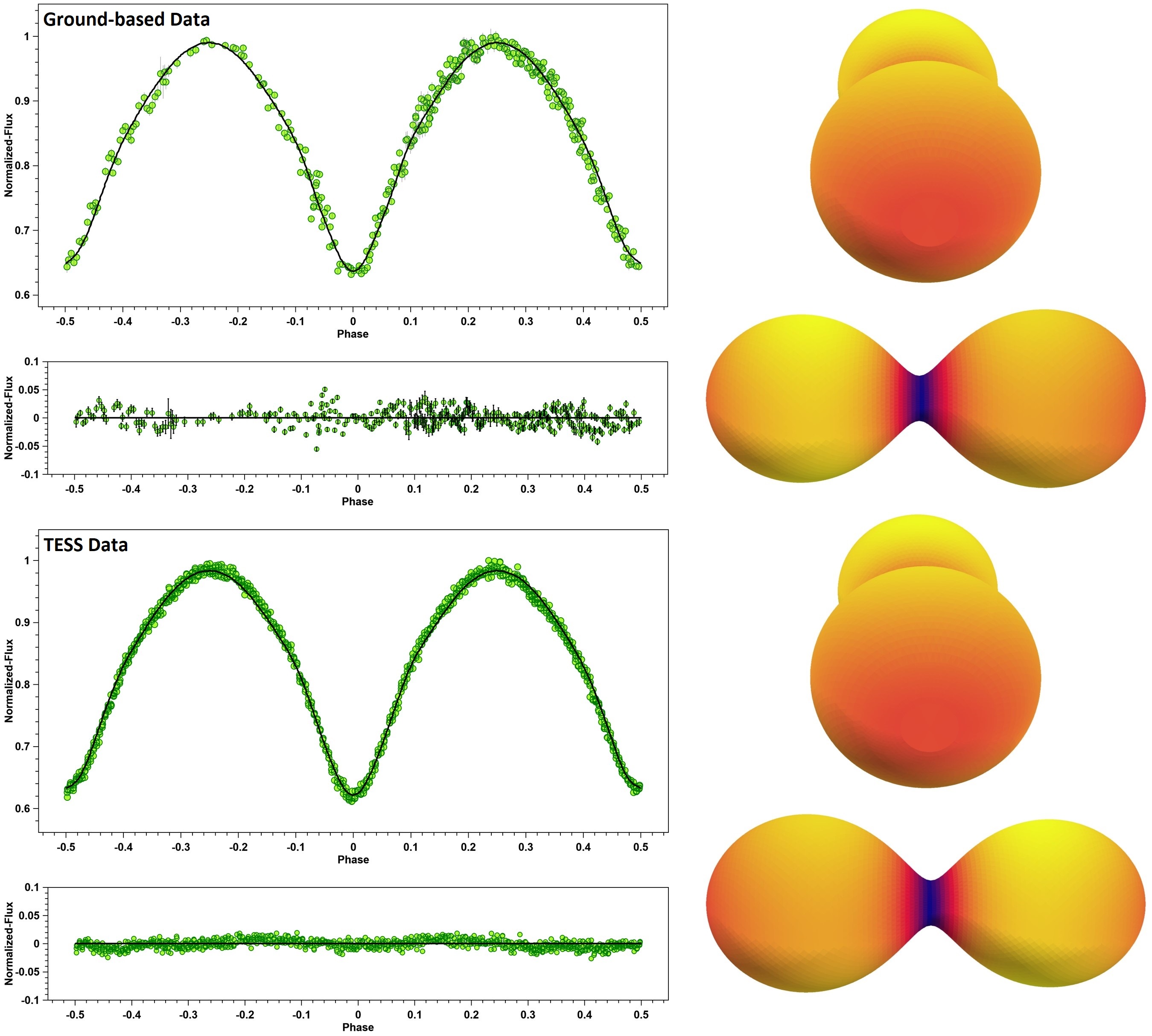}
\caption{Observational (TESS and ground-based) and synthetic light curves, as well as the three-dimensional view in four phases of the V997 Cyg binary system. The data is displayed by the green dots, while the synthetic fit is shown by the black line.}
\label{Fig4}
\end{center}
\end{figure*}

Finally, we employed PHOEBE's optimization tool to improve the output of light curve solutions and derive the final results.
The results of the light curve analysis are shown in Table \ref{tab3}. The available results of other research that are comparable to the results of the current study are also included in Table \ref{tab3}. Figures \ref{Fig3} and \ref{Fig4} include the observational and synthetic light curves and the geometric structure of the systems in four phases. Also, the surface temperature of the stars was displayed in color (Figures \ref{Fig3} and \ref{Fig4}); the brighter the color, the higher the temperature; also, the point of connection between the component stars has a lower temperature due to gravity darkening (\citealt{2016ApJS..227...29P}).

\begin{table*}
\caption{Photometric solutions of the LS Del and V997 Cyg binary systems.}
\centering
\begin{center}
\footnotesize
\begin{tabular}{c | c c c c | c}
 \hline
 \hline
 Parameter & \multicolumn{4}{c|}{LS Del} & V997 Cyg\\
& This study & \cite{1996AA...311..523M} & \cite{1999AJ....118..515L} & \cite{2011MNRAS.412.1787D} & \\
\hline
$T_{1}$ (K) & 6021(37) &5704&&6250(185)& 6629(38)\\
$T_{2}$ (K) & 5959(35) &5780&&6192(168)& 6569(39)\\
$q=M_2/M_1$ & 0.407(50) &0.562&0.375(10)&0.375(10)& 1.078(58)\\
$i^{\circ}$ & 47.81(75) &48.5&48.5&45.25(2.45)& 70.38(25)\\
$f$ & 0.106(31)  &0.06&0.07&0.09& 0.117(18)\\
$\Omega_1=\Omega_2$ & 2.666(45) &&&6.051(80)& 3.811(32)\\
$l_1/l_{tot}(V)$ & 0.699(32) &&&& 0.494(12)\\
$l_2/l_{tot}(V)$ & 0.301(32) &&&& 0.506(12)\\
$l_1/l_{tot}(TESS)$ & 0.701(32) &&&& \\
$l_2/l_{tot}(TESS)$ & 0.299(32) &&&& \\
$r_{1(mean)}$ & 0.468(11) &0.429&&0.308(9)& 0.385(5)\\
$r_{2(mean)}$ & 0.311(10) &0.328&&0.478(8)& 0.398(5)\\
Phase shift & 0.0001(1) &&&& 0.025(1)\\
\hline
Col.(deg) & 83 &&&& \\
Long.(deg) & 56 &&&& \\
Rad.(deg) & 24 &&&& \\
$T_{spot}/T_{star}$ & 0.86 &&&& \\
Component & Secondary &&&& \\
\hline
\hline
\end{tabular}
\end{center}
\label{tab3}
\end{table*}

\vspace{1.5cm}
\section{Absolute Parameters}
There are various methods for estimating absolute parameters. Some of them can be estimated using empirical or statistical relationships between parameters, while one provides an estimate of the absolute parameters by utilizing Gaia's parallax. Gaia DR3's parallax is accurate enough to be used in computations (\citealt{2021AJ....162...13L}). This method of determining absolute parameters makes use of the light curve solution, obtained orbital period, and observational results.
Thus, the absolute magnitude of the system can be determined using the extinction coefficient $A_V$ from the \cite{2019ApJ...887...93G} study, $V_{max}$, and distance. Next, $M_{V_{1,2}}$ values are computed using $l_{1,2}/l_{tot}$ from the light curve solution. We could then estimate each star's absolute bolometric value using the Bolometric Correction (BC) from the \cite{1996ApJ...469..355F} study.
The radius of a star can be calculated from its luminosity and effective temperature using well-known equations in stellar astrophysics.
Also, when we have $r_{mean_{1,2}}$, $a_1$ and $a_2$ can be calculate. It can be inferred that the light curve analysis is appropriate if the values of $a_1$ and $a_2$ are close to each other. We can employ the average of $a_1$ and $a_2$ to determine $a(R_\odot)$.
Finally, by using the orbital period and the mass ratio, the mass of each star is estimated from Kepler's third law equation. This is the method and process we used to estimate the absolute parameters of both systems which are presented in Table \ref{tab4} under the title Method1.
\\
\\
We also estimated the absolute parameters of the LS Del and V997 Cyg systems using the $M_1-P$ relationship (Model2 in Table \ref{tab5}) from the \cite{2021ApJS..254...10L} study.

\begin{equation}\label{eq5}
M_1=(2.94\pm0.21)P+(0.16\pm0.08)
\end{equation}

Then, using the mass ratio, the mass of the other component can be calculated. We estimated the value of $a(R_\odot)$ using Kepler's third law, and then we used well-known $R=a\times r_{(mean1,2)}$ relation to find the star's radius values. The luminosity can be computed using the temperature and radius of the stars, and the absolute bolometric value is then able to be determined for each star. This solutions are presented in Table 4 under the title Method2.

\begin{table*}
\caption{The absolute parameters estimation.}
\centering
\begin{center}
\footnotesize
\begin{tabular}{c | c c c c | c c c c}
 \hline
 \hline
 & \multicolumn{4}{c|}{LS Del} & \multicolumn{4}{c}{V997 Cyg}\\
Parameter &\multicolumn{2}{c}{Method 1} & \multicolumn{2}{c|}{Method 2} & \multicolumn{2}{c}{Method 1} & \multicolumn{2}{c}{Method 2} \\
 & Star1 & Star2 & Star1 & Star2 & Star1 & Star2 & Star1 & Star2\\
\hline
$M(M_\odot)$ & 1.768(114) & 0.719(46) & 1.230(156) & 0.501(133) & 2.255(276) & 2.431(298) & 1.507(176) & 1.625(286) \\
$R(R_\odot)$ & 1.352(29) & 0.909(9) & 1.205(193) & 0.801(136) & 1.595(86) & 1.683(90) & 1.409(283) & 1.456(292) \\
$L(L_\odot)$ & 2.154(42) & 0.934(39) & 1.720(652) & 0.729(292) & 4.406(382) & 4.734(394) & 3.454(1.639) & 3.559(1.693) \\
$M_{bol}(mag.)$ & 3.907(21) & 4.814(46) & 4.141(349) & 5.074(366) & 3.310(92) & 3.052(93) & 3.384(422) & 3.352(422)\\
$log(g)(cgs)$ & 4.424(10) & 4.378(37) & 4.366(77) & 4.330(34) & 4.386(100) & 4.372(101) & 4.318(111) & 4.322(88)\\
$a(R_\odot)$ & \multicolumn{2}{c}{2.906(62)} & \multicolumn{2}{c|}{2.575(344)} & \multicolumn{2}{c}{4.186(172)} & \multicolumn{2}{c}{3.659(679)}\\
\hline
\hline
\end{tabular}
\end{center}
\label{tab4}
\end{table*}

\vspace{1.5cm}
\section{Discussion and Conclusion}
Photometric observations were conducted with a $V$ filter at two French observatories for the LS Del and V997 Cyg systems. Following the standard method, data reduction processes were carried out, and light curves were prepared for analysis. Additionally, TESS data was used for both binary systems in this work. The following conclusions can be derived based on the results of our investigation:

A) Light curve analyses were performed using PHOEBE Python code. We have performed the light curve solution for the LS Del and V997 Cyg systems, and for LS Del it is necessary to add a cold starspot. The obtained mass ratio is in good agreement with other studies (Table \ref{tab3}). The effective temperature difference between the two stars in the LS Del system is 62 K, and in the V997 Cyg system, it is 60 K. According to the temperature obtained in the light curve solutions, it is possible to obtain the spectral category of each star using studies \cite{2000asqu.book.....C} and \cite{2018MNRAS.479.5491E}: For the LS Del system, star1 is G0 and star2 is G1, while both stars of V997 Cyg are F3.
\\
\\
B) As explained in Section 5, we have used two methods to estimate the absolute parameters: Method 1 is based on the Gaia DR3 parallax, and Method 2 uses a relationship between mass and orbital period.

The main method we have chosen for this study is to use Gaia DR3 parallax. The accuracy of this method depends on some parameters, especially Gaia DR3 parallax, $A_V$, $V_{max}$, and light curve solution results.
We calculated $A_V$ using the dust-maps Python package of \cite{2019ApJ...887...93G}. The results show LS Del and V997 Cyg have $A_V$ of 0.061(1) and 0.405(1), respectively.
According to the $A_V$, we can trust estimating absolute parameters using Gaia DR3 parallax for LS Del (Method 1). However, for V997 Cyg with a slightly inappropriate value, we preferred to use Method 2 (\citealt{2024PASP..136b4201P}).
\\
\\
C) We gathered all the times of minima that could be found in the literature and also extracted minima from our observations and space-based data. Based on reference ephemeris, we calculated epochs and O-Cs and presented new ephemeris for each system. The O-C diagrams for both systems show a parabolic fit with an upward trend. Also, we calculated orbital periods increasing rate for each system, which shows less massive star transfer mass to more massive one. The following equation (\ref{eq6}) was used to calculate the mass transferred quantity for the orbital period change:

\begin{equation}\label{eq6}
\frac{\Delta P}{P}=3\Delta M\frac{M_1-M_2}{M_1M_2}
\end{equation}

\begin{figure*}
\begin{center}
\includegraphics[width=\textwidth]{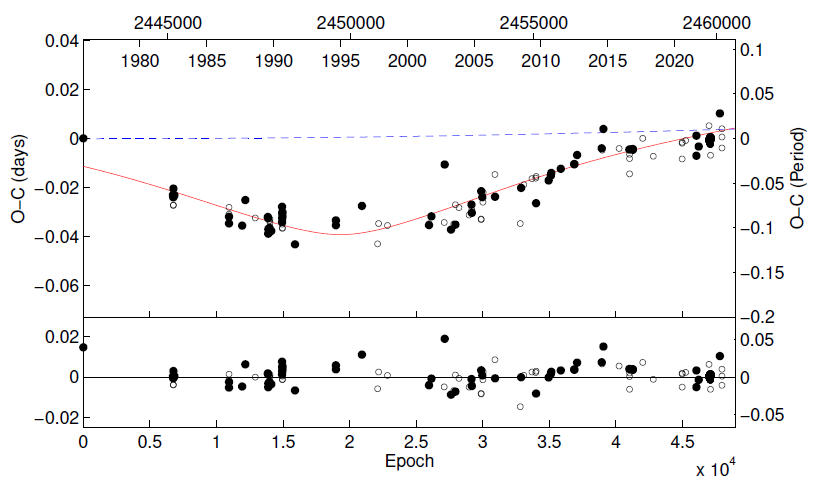}
\caption{O-C diagram considering the possibility of the presence of a third body in the LS Del system.}
\label{Fig5}
\end{center}
\end{figure*}

Taking into account the system elements and $M_1$ and $M_2$ from Table \ref{tab4}, we obtain a transferred mass rate for the LS Del system as $dM/dt=-1.96\times10^{-7}$ $M_\odot$ $yr^{-1}$, and $dM/dt=-3.83\times10^{-7}$ $M_\odot$ $yr^{-1}$ for V997 Cyg.
\\
\\
The main causes of orbital period variation in binary stars include apsidal motion, magnetic activity, the third-body effect, mass, and angular momentum transfer and loss (\citealt{2024NewA..10502112S}). The residual panel in Figure \ref{Fig1} for LS Del shows the possibility of a sinusoidal trend. This becomes significant when the TESS information indicates a possible candidate for the third body in the LS Del system. Therefore, we performed a fit based on the trend in the residual and the presence of the third body (Figure \ref{Fig5}). Also, we calculated the possible third body parameters (Table \ref{tab5}). As can be seen in the residual of Figure \ref{Fig5}, even another specific trend can be understood. It suggests long-term spectroscopic observations of the LS Del system need to be considered in the future.

\begin{figure*}
\begin{center}
\includegraphics[width=\textwidth]{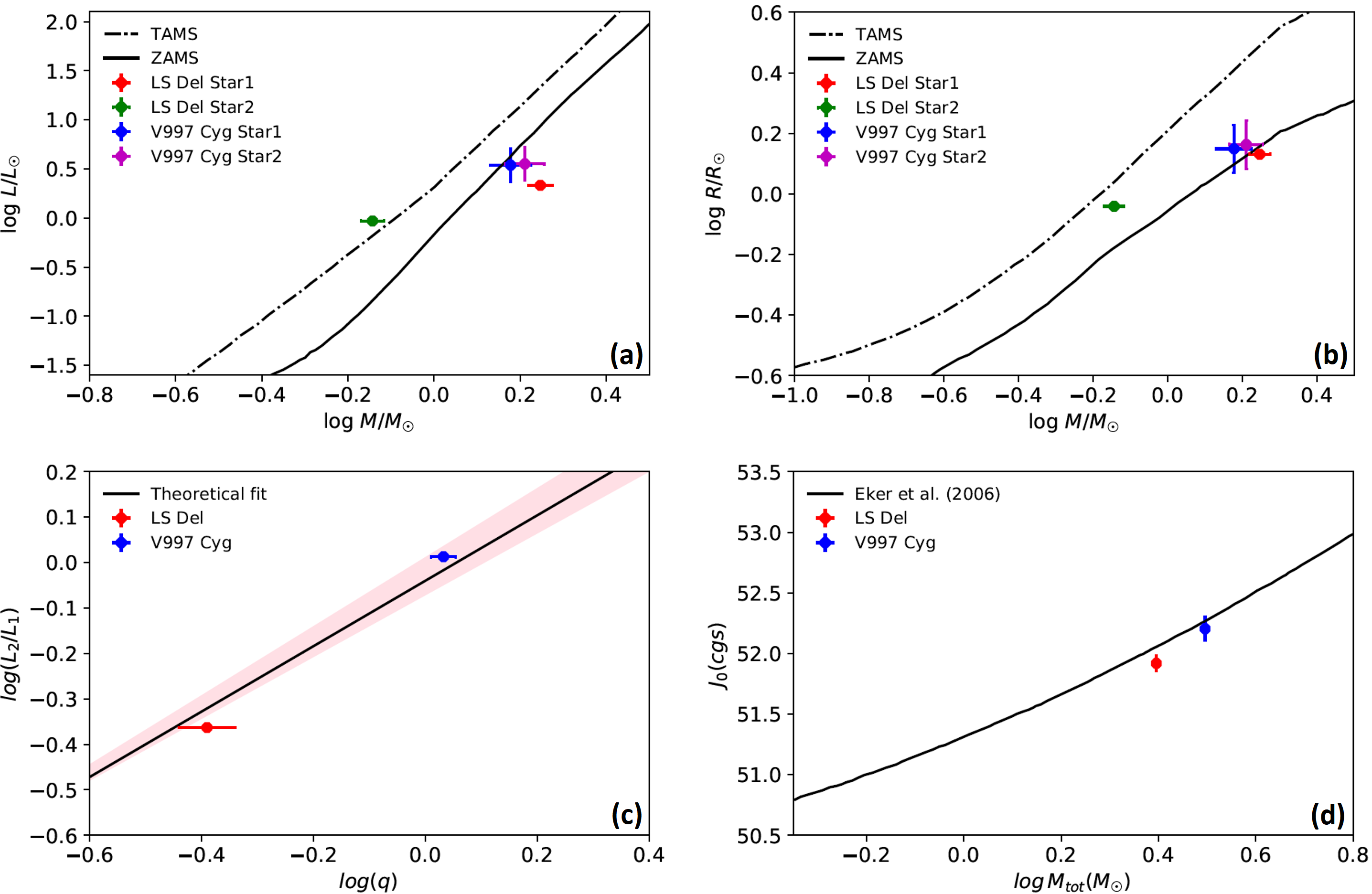}
\caption{The $M-L$, $M-R$, $q-L_{ratio}$, and $M_{tot}-logJ_0$ diagrams.}
\label{Fig6}
\end{center}
\end{figure*}

The results of TESS sector 55, available in the MAST portal and published in PDF format, consider the third object as a planet candidate. TESS reported the temperature of the third object to be $T_{eq}=3170(122)$ K and $T_p=3818(468)$ K. It seems this third body with the mass of $0.29M_{\odot}$ estimated by this study, can be assumed to be an M-type dwarf star. The temperature reported by TESS and the third body mass can be consistent with each other, according to studies \cite{2013ApJS..208....9P}, \cite{2020AA...642A.115C} and \cite{2022Mamajek}\footnote{\url{https://www.pas.rochester.edu/~emamajek/EEM_dwarf_UBVIJHK_colors_Teff.txt}}. As mentioned, we suggested that the fourth body in this system should be investigated in the future.
\\
\\
D) The log-scale Mass-Luminosity ($M-L$) and Mass-Radius ($M-R$) diagrams (Figure \ref{Fig6}a,b) illustrate the target systems' evolutionary stage based on the estimated absolute parameters. Figure \ref{Fig6}a,b shows the positions of the primary and secondary components as well as the theoretical Zero-Age Main Sequence (ZAMS) and Terminal-Age Main Sequence (TAMS) lines. Figure \ref{Fig6}c displays the systems' positions on the $q-L_{ratio}$ relationship, which is in good agreement with the model provided by \cite{2024RAA....24a5002P}.

We calculated the orbital angular momentum ($J_0$) of the systems based on the Equation \ref{eq7} presented by the \cite{2006MNRAS.373.1483E} study.

\begin{equation}\label{eq7}
J_0=\frac{q}{(1+q)^2} \sqrt[3] {\frac{G^2}{2\pi}M^5P}
\end{equation}

The results show that $logJ_0$ for LS Del and V997 Cyg is 51.919(65) and 52.204(98), respectively. The parabolic boundary in Figure \ref{Fig6}d is related to \cite{2024RAA....24a5002P} study, which shows that both of our target systems are in the region of contact binary systems.

\begin{table}
\caption{Parameters and their errors obtained from O-C analysis for LS Del, considering the third body probability.}
\centering
\begin{center}
\footnotesize
\begin{tabular}{c c}
 \hline
 \hline
Parameter & Result\\
\hline
$T_0(BJD_{TDB})$ & 2442687.4183\\
$P_{orb}$(day) & 0.3638404\\
$Q$(day) & $(0.0166\pm0.0003)\times 10^{-10}$\\
$a_{12}sini_3$(AB) & $6.8917\pm0.4248$\\
$A_s$(day) & $0.03979\pm0.00440$\\
$e'$ & $0.7966\pm0.2157$\\
$\omega'$ & $271.2\pm7.7$\\
$T_3(BJD_{TDB})$ & $3027714\pm981$\\
$P_3$(yr) & $263.7\pm0.3$\\
$f(M_3)(M_{\odot})$ & $0.0047\pm0.0003$\\
$(M_3)(M_{\odot})$ for $i=90^{\circ}$ & $0.29$\\
\hline
\hline
\end{tabular}
\end{center}
\label{tab5}
\end{table}

\clearpage
\section*{Acknowledgements}
This manuscript was prepared by the BSN project (\url{https://bsnp.info/}). Project supported by the Scientific Research Projects Coordination Unit of Erciyes University (project number FBA-2022-11737). We have made use of data from the European Space Agency (ESA) mission Gaia (\url{http://www.cosmos.esa.int/gaia}), processed by the Gaia Data Processing and Analysis Consortium (DPAC).
This work includes data from the TESS mission observations. Funding for the TESS mission is provided by the NASA Explorer Program. We thank Ehsan Paki for his comments on the orbital period variation section.

\vspace{1.5cm}
\section*{ORCID iDs}
\noindent Atila Poro: 0000-0002-0196-9732\\
Mehmet Tanriver: 0000-0002-3263-9680\\
Elham Sarvari: 0009-0006-1033-5885\\
Shayan Zavvarei: 0009-0002-2937-9053\\
Hossein Azarara: 0009-0003-2631-6329\\
Laurent Corp: 0009-0003-8727-410X\\
Sabrina Baudart: 0009-0004-8426-4114\\
Asma Ababafi: 0009-0004-8579-7692\\
Nazanin Kahali Poor: 0009-0007-5785-7303\\
Fariba Zare: 0009-0005-4980-0273\\
Ahmet Bulut: 0000-0002-7215-926X\\
Ahmet Keskin: 0000-0002-9314-0648\\

\vspace{1.5cm}
\section*{Data availability}
Ground-based data will be made available on request.

\vspace{1.5cm}
\section*{Appendix}
The appendix tables listed the extracted times of minima for LS Del and V997 Cyg using space-based data. They displayed the times of minima in the first column, the epochs in the second column, the O-C values in the third column, and the references in the fourth column. The minimum time error in all appendix tables is 0.0001. All times of minima in the appendix tables are in $BJD_{TDB}$.

\clearpage
\begin{table*}
\caption{Extracted times of minima of the LS Del binary system from TESS data.}
\renewcommand{\arraystretch}{0.7}
\centering
\begin{center}
\footnotesize
\begin{tabular}{c c c c c c c c c c c c}
 \hline
 \hline
Min. & Epoch & O-C & Ref. & Min. & Epoch & O-C & Ref. & Min. & Epoch & O-C & Ref.\\
\hline
2459797.1937	&	47025.5	&	-0.0015	&	TESS	&	2459805.9255	&	47049.5	&	-0.0020	&	TESS	&	2459815.7488	&	47076.5	&	-0.0023	&	TESS	\\
2459797.3763	&	47026	&	-0.0009	&	TESS	&	2459806.1087	&	47050	&	-0.0006	&	TESS	&	2459815.9326	&	47077	&	-0.0004	&	TESS	\\
2459797.5574	&	47026.5	&	-0.0017	&	TESS	&	2459806.2893	&	47050.5	&	-0.0020	&	TESS	&	2459816.1128	&	47077.5	&	-0.0022	&	TESS	\\
2459797.7402	&	47027	&	-0.0008	&	TESS	&	2459806.4726	&	47051	&	-0.0005	&	TESS	&	2459816.2962	&	47078	&	-0.0007	&	TESS	\\
2459797.9214	&	47027.5	&	-0.0015	&	TESS	&	2459806.6530	&	47051.5	&	-0.0022	&	TESS	&	2459816.4764	&	47078.5	&	-0.0024	&	TESS	\\
2459798.1042	&	47028	&	-0.0006	&	TESS	&	2459806.8365	&	47052	&	-0.0006	&	TESS	&	2459816.6601	&	47079	&	-0.0006	&	TESS	\\
2459798.2850	&	47028.5	&	-0.0017	&	TESS	&	2459807.0171	&	47052.5	&	-0.0019	&	TESS	&	2459816.8405	&	47079.5	&	-0.0022	&	TESS	\\
\hline
\hline
\end{tabular}
\end{center}
\label{tabA1}
\end{table*}

\begin{table*}
\caption{Extracted times of minima of the V997 Cyg binary system from space-based data.}
\renewcommand{\arraystretch}{0.7}
\centering
\begin{center}
\footnotesize
\begin{tabular}{c c c c c c c c c c c c}
 \hline
 \hline
Min. & Epoch & O-C & Ref. & Min. & Epoch & O-C & Ref. & Min. & Epoch & O-C & Ref.\\
\hline
2453170.7566	&	-4997	&	-0.0023	&	TrES	&	2453239.9493	&	-4846	&	-0.0018	&	TrES	&	2454638.6858	&	-1793.5	&	-0.0001	&	SWASP	\\
2453171.9015	&	-4994.5	&	-0.0030	&	TrES	&	2453240.8653	&	-4844	&	-0.0022	&	TrES	&	2454644.6422	&	-1780.5	&	-0.0006	&	SWASP	\\
2453172.8180	&	-4992.5	&	-0.0029	&	TrES	&	2453241.7810	&	-4842	&	-0.0029	&	TrES	&	2454645.5583	&	-1778.5	&	-0.0010	&	SWASP	\\
2453173.9643	&	-4990	&	-0.0022	&	TrES	&	2453242.6975	&	-4840	&	-0.0029	&	TrES	&	2454646.7041	&	-1776	&	-0.0008	&	SWASP	\\
2453174.8797	&	-4988	&	-0.0033	&	TrES	&	2453242.9265	&	-4839.5	&	-0.0031	&	TrES	&	2454647.6203	&	-1774	&	-0.0011	&	SWASP	\\
2453175.7975	&	-4986	&	-0.0019	&	TrES	&	2453243.8440	&	-4837.5	&	-0.0020	&	TrES	&	2454662.5127	&	-1741.5	&	-0.0010	&	SWASP	\\
2453176.7130	&	-4984	&	-0.0029	&	TrES	&	2453244.7604	&	-4835.5	&	-0.0021	&	TrES	&	2454668.4694	&	-1728.5	&	-0.0012	&	SWASP	\\
\hline
\hline
\end{tabular}
\end{center}
\label{tabA2}
\end{table*}

\begin{table*}
\caption{Continued.}
\renewcommand{\arraystretch}{0.7}
\centering
\begin{center}
\footnotesize
\begin{tabular}{c c c c c c c c c c c c}
 \hline
 \hline
Min. & Epoch & O-C & Ref. & Min. & Epoch & O-C & Ref. & Min. & Epoch & O-C & Ref.\\
\hline
2455731.7847	&	592	&	0.0006	&	QES	&	2458689.8658	&	7047.5	&	0.0038	&	TESS	&	2458698.5724	&	7066.5	&	0.0041	&	TESS	\\
2455732.7002	&	594	&	-0.0003	&	QES	&	2458689.8658	&	7047.5	&	0.0038	&	TESS	&	2458698.5724	&	7066.5	&	0.0041	&	TESS	\\
2455733.8467	&	596.5	&	0.0006	&	QES	&	2458690.0948	&	7048	&	0.0037	&	TESS	&	2458698.8009	&	7067	&	0.0035	&	TESS	\\
2455734.7625	&	598.5	&	-0.0001	&	QES	&	2458690.0948	&	7048	&	0.0037	&	TESS	&	2458698.8011	&	7067	&	0.0037	&	TESS	\\
2455735.6796	&	600.5	&	0.0006	&	QES	&	2458690.3238	&	7048.5	&	0.0035	&	TESS	&	2458699.0307	&	7067.5	&	0.0042	&	TESS	\\
2455735.9089	&	601	&	0.0009	&	QES	&	2458690.3238	&	7048.5	&	0.0035	&	TESS	&	2458699.0307	&	7067.5	&	0.0042	&	TESS	\\
2455736.8251	&	603	&	0.0006	&	QES	&	2458690.5532	&	7049	&	0.0039	&	TESS	&	2458699.2600	&	7068	&	0.0044	&	TESS	\\
\hline
\hline
\end{tabular}
\end{center}
\label{tabA3}
\end{table*}

\begin{table*}
\caption{Continued.}
\renewcommand{\arraystretch}{0.7}
\centering
\begin{center}
\footnotesize
\begin{tabular}{c c c c c c c c c c c c}
 \hline
 \hline
Min. & Epoch & O-C & Ref. & Min. & Epoch & O-C & Ref. & Min. & Epoch & O-C & Ref.\\
\hline
2458706.3625	&	7083.5	&	0.0044	&	TESS	&	2458719.4224	&	7112	&	0.0048	&	TESS	&	2458736.3769	&	7149	&	0.0050	&	TESS	\\
2458706.3626	&	7083.5	&	0.0045	&	TESS	&	2458719.6521	&	7112.5	&	0.0055	&	TESS	&	2458736.6065	&	7149.5	&	0.0054	&	TESS	\\
2458706.5913	&	7084	&	0.0041	&	TESS	&	2458719.8811	&	7113	&	0.0053	&	TESS	&	2458736.8352	&	7150	&	0.0051	&	TESS	\\
2458706.5913	&	7084	&	0.0041	&	TESS	&	2458720.1096	&	7113.5	&	0.0047	&	TESS	&	2458737.0646	&	7150.5	&	0.0054	&	TESS	\\
2458706.8206	&	7084.5	&	0.0043	&	TESS	&	2458720.3389	&	7114	&	0.0049	&	TESS	&	2458737.2932	&	7151	&	0.0048	&	TESS	\\
2458706.8206	&	7084.5	&	0.0043	&	TESS	&	2458720.5686	&	7114.5	&	0.0055	&	TESS	&	2458738.8994	&	7154.5	&	0.0072	&	TESS	\\
2458707.0498	&	7085	&	0.0043	&	TESS	&	2458720.7970	&	7115	&	0.0047	&	TESS	&	2458739.1287	&	7155	&	0.0074	&	TESS	\\
\hline
\hline
\end{tabular}
\end{center}
\label{tabA4}
\end{table*}

\begin{table*}
\caption{Continued.}
\renewcommand{\arraystretch}{0.7}
\centering
\begin{center}
\footnotesize
\begin{tabular}{c c c c c c c c c c c c}
 \hline
 \hline
Min. & Epoch & O-C & Ref. & Min. & Epoch & O-C & Ref. & Min. & Epoch & O-C & Ref.\\
\hline
2458754.9374	&	7189.5	&	0.0073	&	TESS	&	2459423.4912	&	8648.5	&	0.0094	&	TESS	&	2459431.5101	&	8666	&	0.0094	&	TESS	\\
2458755.1660	&	7190	&	0.0068	&	TESS	&	2459423.7195	&	8649	&	0.0085	&	TESS	&	2459431.7388	&	8666.5	&	0.0090	&	TESS	\\
2458755.3955	&	7190.5	&	0.0071	&	TESS	&	2459423.7197	&	8649	&	0.0087	&	TESS	&	2459431.7388	&	8666.5	&	0.0090	&	TESS	\\
2458755.6245	&	7191	&	0.0071	&	TESS	&	2459423.9488	&	8649.5	&	0.0087	&	TESS	&	2459431.9679	&	8667	&	0.0089	&	TESS	\\
2458755.8542	&	7191.5	&	0.0077	&	TESS	&	2459423.9488	&	8649.5	&	0.0087	&	TESS	&	2459431.9679	&	8667	&	0.0089	&	TESS	\\
2458756.0824	&	7192	&	0.0068	&	TESS	&	2459424.1777	&	8650	&	0.0086	&	TESS	&	2459432.1967	&	8667.5	&	0.0086	&	TESS	\\
2458756.3119	&	7192.5	&	0.0071	&	TESS	&	2459424.1778	&	8650	&	0.0087	&	TESS	&	2459432.1967	&	8667.5	&	0.0086	&	TESS	\\
\hline
\hline
\end{tabular}
\end{center}
\label{tabA5}
\end{table*}

\begin{table*}
\caption{Continued.}
\renewcommand{\arraystretch}{0.7}
\centering
\begin{center}
\footnotesize
\begin{tabular}{c c c c c c c c c c c c}
 \hline
 \hline
Min. & Epoch & O-C & Ref. & Min. & Epoch & O-C & Ref. & Min. & Epoch & O-C & Ref.\\
\hline
2459440.4452	&	8685.5	&	0.0090	&	TESS	&	2459773.5766	&	9412.5	&	0.0101	&	TESS	&	2459790.3013	&	9449	&	0.0096	&	TESS	\\
2459440.6740	&	8686	&	0.0087	&	TESS	&	2459773.8055	&	9413	&	0.0099	&	TESS	&	2459790.5311	&	9449.5	&	0.0103	&	TESS	\\
2459440.6740	&	8686	&	0.0087	&	TESS	&	2459774.0348	&	9413.5	&	0.0101	&	TESS	&	2459790.7601	&	9450	&	0.0102	&	TESS	\\
2459440.9038	&	8686.5	&	0.0094	&	TESS	&	2459774.2635	&	9414	&	0.0097	&	TESS	&	2459790.9893	&	9450.5	&	0.0102	&	TESS	\\
2459440.9038	&	8686.5	&	0.0094	&	TESS	&	2459774.4930	&	9414.5	&	0.0101	&	TESS	&	2459791.2184	&	9451	&	0.0103	&	TESS	\\
2459441.1327	&	8687	&	0.0092	&	TESS	&	2459774.7218	&	9415	&	0.0098	&	TESS	&	2459791.4477	&	9451.5	&	0.0104	&	TESS	\\
2459441.1327	&	8687	&	0.0092	&	TESS	&	2459774.9516	&	9415.5	&	0.0105	&	TESS	&	2459791.6768	&	9452	&	0.0104	&	TESS	\\
\hline
\hline
\end{tabular}
\end{center}
\label{tabA6}
\end{table*}

\begin{table*}
\caption{Continued.}
\renewcommand{\arraystretch}{0.7}
\centering
\begin{center}
\footnotesize
\begin{tabular}{c c c c c c c c c c c c}
 \hline
 \hline
Min. & Epoch & O-C & Ref. & Min. & Epoch & O-C & Ref. & Min. & Epoch & O-C & Ref.\\
\hline
2459801.9868	&	9474.5	&	0.0103	&	TESS	&	2459809.0894	&	9490	&	0.0104	&	TESS	&	2459817.3373	&	9508	&	0.0103	&	TESS	\\
2459801.9868	&	9474.5	&	0.0103	&	TESS	&	2459809.3179	&	9490.5	&	0.0098	&	TESS	&	2459817.3373	&	9508	&	0.0103	&	TESS	\\
2459802.2160	&	9475	&	0.0104	&	TESS	&	2459809.3179	&	9490.5	&	0.0098	&	TESS	&	2459817.5658	&	9508.5	&	0.0096	&	TESS	\\
2459802.2162	&	9475	&	0.0106	&	TESS	&	2459809.5474	&	9491	&	0.0102	&	TESS	&	2459817.5660	&	9508.5	&	0.0098	&	TESS	\\
2459802.4448	&	9475.5	&	0.0101	&	TESS	&	2459809.5474	&	9491	&	0.0102	&	TESS	&	2459817.7953	&	9509	&	0.0100	&	TESS	\\
2459802.4453	&	9475.5	&	0.0106	&	TESS	&	2459809.7764	&	9491.5	&	0.0101	&	TESS	&	2459817.7953	&	9509	&	0.0100	&	TESS	\\
2459802.6743	&	9476	&	0.0104	&	TESS	&	2459809.7764	&	9491.5	&	0.0101	&	TESS	&	2459818.0240	&	9509.5	&	0.0096	&	TESS	\\
\hline
\hline
\end{tabular}
\end{center}
\label{tabA7}
\end{table*}

\clearpage
\bibliography{Ref}{}
\bibliographystyle{aasjournal}

\end{document}